\documentclass[letter,draftclsnofoot,onecolumn,12pt,oneside]{IEEEtran}
\pdfoutput=1
\usepackage{cite}
\usepackage{epsfig,graphics,mathrsfs,amssymb,amsmath,bm,color,yfonts,txfonts,multirow,multicol,slashbox,dblfloatfix}
\usepackage{subfigure}
\usepackage{algorithm2e}

\begin{document}

\title{Outage Analysis of Offloading in Heterogeneous Networks: Composite Fading Channels}
\author{Mirza G. Kibria, Gabriel P. Villardi, Wei-Shun Liao, Kien Nguyen, Kentaro Ishizu and Fumihide Kojima
 \thanks{The authors are with Wireless Systems Laboratory,  Wireless Networks Research Center, National Institute of Information and Communications Technology (NICT), Yokosuka Research Park, Japan 239-0847 (e-mails: $\left\{\text{mirza.kibria, gpvillardi, wsliao, kienng, ishidu, f-kojima}\right\}@\text{nict.go.jp}$). }
}

%

\maketitle

\begin{abstract}

Small cells deployment is one of the most significant long-term strategic policies of the mobile network operators.
In heterogeneous networks (HetNets), small cells serve as offloading spots in the radio access network to offload macro users (MUs) and their associated traffic from congested macrocells. In this paper, we perform analytical analysis and investigate how the radio propagation effects such as multipath and shadowing and small cell base station density affect MUs' offloading to small cell network (SCN). In particular, we exploit composite fading channels in our evaluation when an MU is offloaded to SCN with varying small and macro cell densities in the stochastic HetNets framework. We derive the expressions for outage probability (equivalently success probability) of the MU in macro network and SCN for two different cases, viz.: i) Nakagami-lognormal channel fading; ii) time-shared (combined) shadowed/unshadowed channel fading. We propose efficient approximations for the probability density functions of the channel fading (power) for the above-mentioned fading distributions that do not have closed-form expressions employing Gauss-Hermite integration and finite exponential series, respectively. Finally, the outage probability performance of MU with and without offloading options/services is analyzed for various settings of fading channels.

\end{abstract}

\IEEEpeerreviewmaketitle
\section{Introduction}

\IEEEPARstart{T}{he} evolution of cellular networks over the last few decades has been impressive. Networks planned for voice traffic now accommodate and support high data-traffic loads, and this growth has mainly dependent on a network of large cells or macro cells providing an agreeable evenness of capacity and coverage. The network operators have already started to devise new solutions in order to expand and maximize capacity in high-traffic regions, while the macro cell network (MCN) will still continue to deliver necessary wide-range coverage and support for high-mobility users. An advanced approach is strongly required to support a gain in capacity of the magnitude enough to handle the ever increasing capacity demand, and this approach is set on shorter-range, lower-power small cells placed closer to the users, especially in denser deployments. The needs of high data throughputs and improved coverage for home and office use, small cells have attracted significant interest in the wireless industry.
Small cell deployments are one of the most significant long-term strategic policies of the network operators since small cell architecture delivers not only the required capacity boost but also the flexibility important for immensely localized deployments, and also provides a higher satisfaction quotient from the subscribers' standpoint. Therefore, heterogeneous networks (HetNets), consisting of macro cells overlaid with small cells provide a cost-efficient, flexible and fine-tuned design, and facilitates expansion of existing cellular networks to meet the ever increasing network capacity demand. A further knock-on benefit of small cells is their potential to significantly reduce the network energy consumption if integrated with advanced energy saving techniques.

Traffic offloading has been found to be a satisfyingly and most widely adopted solution where usage of cellular data and users' density are high. In HetNets, small cells serve as offloading spots to offload users and their affiliated traffic from overloaded or congested macrocells. By bringing the radio access network infrastructure closer to the user, small cell network (SCN) has the capability in affording a better link budget, which translates into higher spectrum efficiency and efficient use of network resources to support a targeted capacity increase where needed at a much lower cost. Due to shorter distances between the transmitter-receiver pair, the transmit power required to achieve the same quality of service (QoS) scales down significantly in the small cell scenario. 
For accurate small cell deployments planning and performance evaluations, the impact of channel propagation impairments such as large-scale fading, which arises from shadowing, and small-scale fading, which is due to multipath propagation need to take into account \cite{Yeh, ElSawy}. In a dense small cells deployment with aggressive frequency reuse in neighboring cells, i.e., in an interference-limited environment, co-channel interference should also be considered as a corruptive effect. Both the desired and interfering signals are subject to multipath and shadow fading, and it is necessary to incorporate these effects in assessing the performance of user offloading based wireless systems. 

\subsection{Related Works}

In \cite{Rebecchi} and the references therein, a comprehensive survey of data offloading techniques in cellular networks and main requirements required to integrate data offloading capabilities in mobile networks have been presented. 
In \cite{Kang}, the authors investigated the coexistence of MCN and SCN where small cell base stations (SBSs) are allowed to transmit to their users as long as the aggregated interference remain below a certain threshold. In \cite{Singh1}, the authors analyzed the impacts of user offloading in a multi-tiered HetNets under a flexible received power association model. In \cite{Yun} and the references therein, traditional spectrum access models under offloading to SCN classified under different spectrum access paradigms are presented. In \cite{Chen, Hafeez}, repayment based offloading schemes under the objective of enhancing the energy efficiency are proposed, where the macro users (MUs) are granted SCN resources
in exchange for mutual profit reimbursement. In \cite{Wildemeersch}, the authors investigated the coexistence of SCNs and MCNs under objective of lowering the energy consumption of the MCN by offloading traffic to SCN. Note that in \cite{Mueck}, the authors have stated that the problem of offloading mobile network operator's traffic to small cells is an interesting open research issue, which validates the timeliness and importance of our work.

In \cite{Chen1, Chen2}, the authors proposed an energy-aware data offloading scheme in order to make the offloading procedure more efficient while preserving satisfactory QoS for the offloaded users. In \cite{Suh}, the authors studied two different types of offloading techniques (offloading the MUs to WiFi), namely, i) opportunistic offloading, where offloading takes place when the MU opportunistically meets WiFi access points and ii) delayed WiFi offloading. A threshold-based distributed offloading scheme is also proposed in \cite{Suh} to release the burden on the macro base station (MBS). In \cite{Liu}, the authors proposed traffic offloading from MBS to low power nodes, such as pico stations and explored the energy efficiency gains. The scheme is based on determining the optimal set of pico cells for offloading that improves the energy efficiency. In \cite{Zhuo}, the authors investigated the tradeoff between users's satisfaction and the amount of traffic being offloaded, and proposed a novel incentive framework that encourages the MUs to leverage their delay tolerance for traffic offloading. In \cite{ElSawy1}, the authors quantified offloading and discussed several techniques to offload the users from macro access network to SCN, namely, offloading via small cell deployment, offloading via power control and offloading based on biasing in a Nakagami-$m$ fading environment. However, the offloading scheme discussed above considered idealised fading scenarios and do not consider diverse fading scenarios.

\subsection{Contributions}
Most of the offloading schemes discussed above considered idealised radio propagation models and do not consider diverse fading scenarios, i.e., do not consider realistic fading and shadowing effects. Note that offloading decision is significantly affected by the wireless channel propagation characteristics. In such an environment, the link performance evaluation depends on many channel parameters. To assess the impact of these different parameters on system evaluation metrics, namely, the outage probability, closed-form and tractable expressions are highly desirable.

In this paper, we focused on the problem of analytically modeling the behavior of MUs' offloading to SCN. To the best of our knowledge, this is the first analytical study that elaborates on the performance of these offloading techniques under diverse fading environments. The paper mainly presents three key contributions:

\begin{itemize}
\item Formulation of approximated closed-form expression for probability density function of shadowed Nakagami-$m$ and time-shared (combined) shadowed/unshadowed channel fading channels. We propose efficient approximations for the probability density functions of the channel fading (power) that do not have closed-form expressions employing Gauss-Hermite (for Nakagami-$m$) and finite exponential series (for time-shared shadowed/unshadowed fading). 
\item Analysis on outage probability of MUs with and without offloading served by SCN within the stochastic HetNets framework.
\item Evaluation on impact of SBS density and channel fading parameters on outage performance of MUs. 
\end{itemize}

\begin{figure*}
  \centering
   \includegraphics[scale=.535]{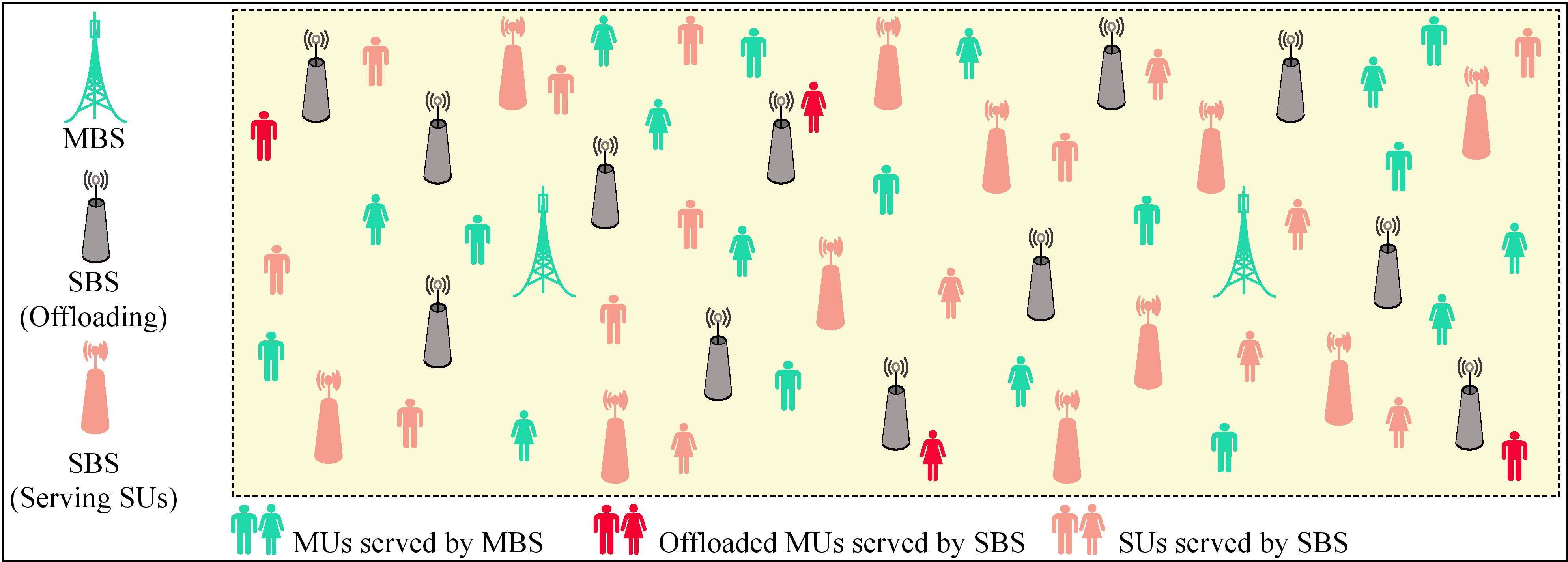}
   \caption{Schematic of the heterogeneous system model. The SBSs belonging to the SCN are classified under two groups, namely licensee SBSs and offloader SBSs. The actual division of the SBSs into these two groups, i.e., sizes of the groups depend on the negotiation/agreement between the MCN network and SCN network.}
   \label{figZ}
\end{figure*}


 The remainder of this paper is organized as follows. In Section \ref{SM} discusses the system model under stochastic geometry considered in this study. Section \ref{DM} presents the fading scenarios we have considered and the approximations for fading (power) probability distribution functions. Section \ref{PM} illustrates analytical analysis for outage probability in both direct and offloading modes. Section \ref{XM} presents the simulation results, and finally, Section \ref{YM} concludes the paper.


\section{System Model }
\label{SM}

A two-tier heterogeneous cellular network is considered, where the tiers designate the base stations (BSs) of different classes, for example, such as those of macro cells, micro cells, metro cells, pico cells or femto cells. The BSs across different tiers may diverge in terms of supported maximum transmitting power, data rates and the spatial intensity of their deployments. We consider that the spatial locations of the MBSs and SBSs are independent and distributed as Poisson point process (PPP) in two-dimensional Euclidean planes. Let us consider that the MBSs and SBSs are distributed as PPP $\Phi_\text{M}\subset\mathbb{D}^2$ of intensity $\lambda_{\text{M}}$ and PPP $\Phi_\text{S}\subset\mathbb{D}^2$ of intensity $\lambda_{\text{S}}$, respectively. All the MBSs in $\Phi_\text{M}\subset\mathbb{D}^2$ and SBSs in $\Phi_\text{S}\subset\mathbb{D}^2$ transmit at identical transmitting powers $P_{\text{M}}$ and $P_{\text{S}}$, respectively. Since HetNets are interference-limited, in this paper, we
ignore the thermal noise in the analysis, as in\cite{Heath}. Therefore, we consider a highly spectrally efficient interference-limited scenario in which the noise component is negligible compared to the co-channel interference \cite{Lee}. Neglecting the noise also makes the analysis more tractable. Thus, the signal to interference plus noise power ratio (SINR) reduces to the signal to interference power ratio (SIR). The MUs have SIR target of $\mu_\text{M}$. Each SBS/MBS serves one user at each time slot, i.e., the considered network becomes a time division multiple access (TDMA) HetNet.

We assume a general power-law path loss model in which the signal power decays at the rate $r_m^{-\eta}$ with the distance $r_m$,
where $\eta>2$  is the path-loss exponent. The fading (power) between the BS at a distance of $r_m$ from the tagged MU is denoted by $h_m$, which can be either shadowed Nakagami-$m$ or time-shared shadowed/unshadowed fading. Under the long term evolution (LTE) context, we limit our analytical study to a specific transmission time interval (TTI) or slot and we assume i.i.d block fading over TTIs. Hence, the received power at any MU at the center who is at a distance of $r_m$ from the BS is given by  $P_\text{x}h_mr_m^{-\eta},\hspace{1mm}\text{x}\in\{\rm{M},\rm{S}\}$. We assume that the MCN and the SCN operate over orthogonal frequencies. Thus, there is no inter-network interference between MCN and SCN. Therefore, for a generic user located at the center, the aggregated interference power comes from all the MBSs or SBSs (depending on the modes of operation) operating on the same channel. It should be noted that many MUs likely to be competing for wireless services within a single cell, and the MBS selects a particular MU that will be served in a designated time slot with probability $p_{\text{s}}^{\text{m}}$. 
A schematic of the considered system model and offloading strategy is provided in Fig.~\ref{figZ}, where the SCN offers offloading service to the MCN, and in return, the SCN is rewarded for its cooperation. The offered incentive is in terms of number of licences to operate in the spectrum originally owned by the MCN. A division between the fraction of SBSs providing offloading services and those who obtain the licenses to operate in MCN's spectrum needs to be determined.

\section{Diverse fading scenarios and channel fading (power) PDF approximation }
\label{DM}

\subsection{Shadowed Nakagami-$m$ Fading}

When signal propagates through the wireless medium, it undergoes deleterious effects mainly characterized by path-loss, multipath fading and shadowing. In such composite fading environment, the receiver reacts to the composite multipath-shadowed signal, rather than averaging out the envelope fading due to multipath. This fading scenario is very often observed in congested downtowns areas with slow-moving vehicles and pedestrians. This type of composite fading can also be observed in land-mobile satellite systems subject to urban and/or vegetative shadowing. In this analysis, we study the composite lognormal/gamma probability density function (PDF) introduced by \cite{Stuber}, which is obtained by averaging the instantaneous Nakagami-$m$ fading amplitude over the PDF of Lognormal fading amplitude as follows

\begin{equation}
\begin{aligned}
\label{NLPDF}
f_{z}(z)&=\hspace{-2mm}\int_0^{\infty}f_z(z\hspace{1mm}|\hspace{1mm}\sigma)f_\sigma(\sigma)\text{d}\sigma\\
&=\hspace{-2mm}\int \limits_0^{\infty}\frac{z^{2m-1}\exp\left(-\frac{mz^2}{\sigma}\right)}{\Gamma(m)\sigma^m}\frac{2m^m}{\sqrt{2\pi}\zeta\sigma}\exp\left(-\frac{(\ln \sigma-\mu)^2}{2\zeta^2}\right)d\sigma.
\end{aligned}
\end{equation}
where, $z$ is a Nakagami-$m$ random variable, representing fast fading, whose PDF is given by
$
f_z(z\hspace{1mm}|\hspace{1mm}\sigma)=\frac{2m^mz^{2m-1}}{\Gamma(m)\sigma^m}\exp\left(-\frac{mz^2}{\sigma}\right).
$
Here, $\sigma=\mathbb{E}[z^2]$ is the average signal power and $m$ is the Nakagami fading parameter that controls the severity of the amplitude fading. The value $m=1$ results in the wide-spread Rayleigh-fading model, while values of $m<1$ correspond to channel fading more severe than Rayleigh fading and values of $m>1$ correspond to channel fading less severe than Rayleigh fading. 
Furthermore, $f_\sigma(\sigma)=\frac{1}{\sqrt{2\pi}\zeta \sigma}\exp\left(-\frac{(\ln(\sigma)-\mu)^2}{2\zeta^2}\right)$ is the PDF of slow-varying local mean $\sigma$, which is a lognormal random variable that corresponds to shadowing. The parameters $\mu=\frac{\ln(10)}{10}\mu_G$ and $\zeta=\frac{\ln(10)}{10}\zeta_G$, where $\mu_G$ and $\zeta_G$ are the mean and standard deviation of the associated Gaussian process of the shadowing and associated normal process related to the lognormal shadowing, respectively.

 
 Note that the PDF in \eqref{NLPDF} and its corresponding cumulative density function (CDF) are not expressible in closed-from, and as a result, the composite fading distribution in \eqref{NLPDF} does not lend itself to performance analysis readily. Therefore, in the following, we attempt to derive a simpler approximation for distribution of the composite Nakagami-lognormal fading.
With the substitution $t=\frac{(\ln \sigma-\mu)}{\sqrt{2}\zeta}$ (equivalently, $\sigma = \exp\left(\sqrt{2}\zeta t+\mu\right)$, we can express \eqref{NLPDF} as the following
\begin{equation}
f_{z}(z)=\frac{2m^mz^{2m-1}}{\Gamma(m)\sqrt{\pi}}\int \limits_{-\infty}^{\infty}\exp\left(-t^2\right)\Psi(t)dt,
\end{equation}
with $$\Psi(t)= \exp\left(-\exp\left(-\sqrt{2}\zeta t - \mu\right)mz^2\right)\left(\exp\left(\sqrt{2} \zeta t + \mu\right)\right)^{-m}.$$ Note that the term $\int \limits_{-\infty}^{\infty}\exp\left(-t^2\right)\Psi(t)dt$ is in the form of Gauss-Hermite integration and thus, can be approximated as

\begin{equation}
\int \limits_{-\infty}^{\infty}\exp\left(-t^2\right)\Psi(t)dt\approx \sum_{i=1}^{N}\mathcal{W}(i)\Psi(t_i)
\end{equation}
where $\mathcal{W}(i)$ and $t_i$ are the weights and abscissas ($i$-th root of an $N$-th order Hermite polynomial) of the Gauss-Hermite quadrature. Values of $\mathcal{W}(i)$ and $t_i$ for different $N$ values are listed in \cite{Abramowitz}. Therefore, the approximated PDF of the composite fading can be expressed as 
\begin{equation}
f_{z}(z)\approx\frac{2m^mz^{2m-1}}{\Gamma(m)\sqrt{\pi}}\sum_{i=1}^{N}\mathcal{W}(i)\Psi(t_i).
\end{equation}

Note that the received instantaneous signal power is modulated by $z^2$. In order to find the channel gain (power) distribution, let $h$ be the random variable which is the square of Nakagami-lognormal random variable $z$, ie., $h=z^2/2$ defines the channel fading power, and $f_h(h)$ be its corresponding PDF. Therefore, $f_h(h)$ is the transformed PDF of $f_{z}(z)$, where $h=z^2/2$, i.e.,
\begin{equation}
\label{APDF}
f_h(h)\approx\text{Transformed-PDF}\left[h=z^2, z\sim f_{z}(z)\right],
\end{equation}
which is given by \eqref{PDFFH} with $[\alpha_1,\cdots, \alpha_{N/2-1}]=[0.000733446,\hspace{1mm}0.019126,\hspace{1mm}0.135462,\hspace{1mm}0.344663]$ and $[\beta_1,\cdots, \beta_{N/2-1}]=[3.58178, \hspace{1mm}2.48435, \hspace{1mm}1.46597, \hspace{1mm}0.484934]$ for $N=10$.
In order to check the closeness of this approximation to the original integral-form, we put $m=1$ and $\sigma=0$ in \eqref{PDFFH}, and we know that \eqref{PDFFH} should reproduce an exponential PDF as the channel fading power has exponential distribution when the fading amplitude is Rayleigh ($m=1$) distributed. We have found that \eqref{PDFFH} reproduces the exponential PDF $0.9997\exp\left(-h\right)$ for $m=1$, which is very close to $\exp\left(-h\right)$.

\begin{equation}
\begin{aligned}
\label{PDFFH}
f_h(h)&\approx \frac{m^m h^{m-1} }{\Gamma[m]} \left(\sum_{i=1}^{N/2-1}\alpha_i \exp\left(-mh\exp\left(\beta_i \sigma -\mu\right) \right)\left(\exp\left(-\beta_i \sigma +\mu \right)\right)^{-m}\right.\\
& \left.+\sum_{i=1}^{N/2-1}\alpha_i \exp\left(-mh\exp\left(-\beta_i \sigma -\mu\right) \right)\left(\exp\left(\beta_i \sigma +\mu \right)\right)^{-m} \right) 
\end{aligned}
\end{equation}

\subsection{Time-Shared Shadowed/Unshadowed Fading }


From the land-mobile satellite channel characterization experiments, it has been observed that the overall fading process is a convex combination of unshadowed multipath fading and a composite multipath/shadowed fading \cite{Lutz,Barts}. According to this fading environment, when there is no shadowing, the fading follows a Rician (Nakagami-$n$) distribution. On the contrary, when shadowing is present, the assumption is that no direct line-of-sight paths exist and the received signal power is assumed to be exponentially/lognormally distributed  \cite{Hansen}. The convex combination is characterized by the shadowing time-share factor, $\mathcal{T}$, $0\le \mathcal{T}\le1$. The resulting combined PDF is given by \eqref{PDFFHX1},
\begin{equation}
\label{PDFFHX1}
f_z(z)=(1-\mathcal{T})\frac{x}{\psi^2}\text{exp}\left(\frac{-(z^2+\gamma^2)}{2\psi^2}\right)I_0\left(\frac{z\gamma}{\psi^2}\right)+\mathcal{T}\int \limits_0^{\infty}\frac{2z}{\sigma}\exp\left(-\frac{z^2}{\sigma}\right)\frac{1}{\sqrt{2\pi}\zeta\sigma}\exp\left(-\frac{(\ln \sigma-\mu)^2}{2\zeta^2}\right)d\sigma
\end{equation}
with $\mu=\frac{\ln(10)}{10}\mu_G$ and $\zeta=\frac{\ln(10)}{10}\zeta_G$, where $\mu_G$ and $\zeta_G$ are the mean and standard deviation of the associated Gaussian process of the shadowing and associated normal process related to the lognormal shadowing, respectively, during the shadowed fraction of time.

Let us consider that the amplitude of the channel fading has Rician distribution, and $X$ represents the amplitude the Rician distributed channel fading. Then the PDF of random variable $X$ is given by
\begin{equation}
f^{(R)}_{z}(z)=\frac{{z}}{\psi^2}\text{exp}\left(\frac{-({z}^2+\gamma^2)}{2\psi^2}\right)I_0\left(\frac{{z}\gamma}{\psi^2}\right), z\ge0
\end{equation}
where, $I_n(\cdot)$ is the modified Bessel function of the first kind with order $n$. Let $h$ be the random variable which is the square of Rician distributed variable ${z}$, ie., $h={z}^2$ defines the channel fading power. Note that $\gamma^2$ is the power in the specular path while $2\psi^2$ is power in the scattered multipaths.
Now, the PDF of $h$ is given by the Dirac-delta method as
%

\begin{equation}
\label{RICIAN}
\begin{aligned}
f_h(h)&=\int_0^\infty \text{d}z\ \frac{x}{\psi^2}\text{exp}\left(\frac{-(z^2+\gamma^2)}{2\psi^2}\right)I_0\left(\frac{x\gamma}{\psi^2}\right)\delta(h-z^2)\\
&=\int_0^\infty \text{d}z\ \frac{z}{\psi^2}\text{exp}\left(\frac{-(z^2+\gamma^2)}{2\psi^2}\right)I_0\left(\frac{z\gamma}{\psi^2}\right)\frac{\delta(z-\sqrt{h})}{2\sqrt{h}}\\
&=\frac{1}{2\psi^2}\exp\left(\frac{-(h+\gamma^2)}{2\psi^2}\right)I_0\left(\frac{\gamma\sqrt{h}}{\psi^2}\right),\
\end{aligned}
\end{equation}
Note that in the absence of a dominant scatterer, $\gamma^2=0$, \eqref{RICIAN} reduces to the exponential PDF, as should be expected. (This result uses the fact that $I_0(0) = 1$.). Note that $\int_0^\infty f_h(h)\text{d}H=1$, so the PDF is correctly normalized for $\gamma^2=0$. Note that for other values of $\gamma^2$, the PDF is not correctly normalized, thus a normalization factor needs to be incorporated to the PDF.  Now, in order to have a better tractability, we express $f_{h}(h)$ in terms of the Rician factor $K$, where $K=\gamma^2/2\psi^2$ is the ratio of power in specular path to the power in scattered multipaths. We know that Rician  fading is generally described by one more parameter, which is $\Theta=\gamma^2+2\psi^2$, the total power from both the direct path and the multipaths. The power in direct path in terms of $K$ and $\Theta$ is given by $\gamma^2=\frac{K}{1+K}\Theta$, while $\psi^2$ is given by $\psi^2=\frac{\Theta}{2(1+K)}$. Now, substituting the values of $\gamma^2$ and $\psi^2$ in \eqref{RICIAN}, we can express the PDF\footnote{Note that Rician fading channel is also known as Nakagami-$n$ fading channel for which the PDF of channel fading $H$ is given by
$
f^{(R)}_h(h)=\frac{n^2+1}{\Theta}\exp\left(-n^2-\frac{(n^2+1)h}{\Theta}\right)I_0\left(2\sqrt{\frac{n^2(n^2+1)h}{\Theta}}\right)$
with $n^2=K$. The PDF $f_h(h)$ in \eqref{RICIAN1} can be further simplified to the following expression using Hypergeometric function instead of Bessel function as
$f_h(h)\triangleq\frac{\exp\left(-(h+hK+K\Theta)\right)(1+K)_0F_1(;1;-\frac{hK(1+K)}{\Theta})}{\Theta}$
where $_0F_1(;b;c)$ is the confluent Hypergeometric function given by $_0F_1(;b;c)=\sum_{k=0}^\infty\frac{c^k}{(b)_kk!}$. Here, $(b)_k$ denotes the Pochhammer symbol with $(b)_k=\Gamma[b+k]/\Gamma[b].$} of channel fading power as
\begin{equation}
\label{RICIAN1}
\begin{aligned}
f^{(R)}_{h}(h)&=\frac{K+1}{\Theta}\exp\left(-K-\frac{(K+1)h}{\Theta}\right)I_0\left(2\sqrt{\frac{K(K+1)h}{\Theta}}\right)\\
\end{aligned}
\end{equation}
Like \eqref{RICIAN}, the PDF expression in \eqref{RICIAN1} is also not correctly normalized for general values of $K$ other than case $K=0$. A normalization factor $\exp(2K)$ should be multiplied with PDF in \eqref{RICIAN1}. 

Note that the PDF expression in \eqref{RICIAN1} is not in the closed-form because of the Bessel function. One must resort to numerical integration techniques or to approximated expressions in order to perform numerical evaluation. We employ a very efficient and simple approximation to the Bessel function $I_0(x)$. There have been extensive studies on approximating the Bessel function in terms of exponential, trigonometric and polynomial expansions \cite{Gross, Millance}. Based on non-oscillatory behaviour of the modified Bessel functions, a good approximation model given as a sum of pure real finite exponential series is proposed in \cite{Salahat}, for which the term $I_0\left(2\sqrt{\frac{K(K+1)h}{\Theta}}\right)$ in \eqref{RICIAN1} can be approximated\footnote{The normalization factor $\exp(2K)$ is not required for the approximated PDF. It is already correctly normalized for any $K$ value.} as
\begin{equation}
\label{RICIAN2}
I_0\left(2\sqrt{\frac{K(K+1)h}{\Theta}}\right)\approx \sum_{k=1}^{4}\alpha_k\exp\left({2\sqrt{\frac{K(K+1)h}{\Theta/{\beta_i^2}}}}\right),
\end{equation}
where the values of $\alpha$'s and $\beta$'s are listed in Table 1 of \cite{Salahat}. Therefore, the approximated PDF for the power of Rician faded signal is given by
     \begin{equation}
     \label{PDFF}
       \begin{aligned}
f^{(R)}_{h}(h)\approx \frac{K\hspace{-1mm}+\hspace{-1mm}1}{\Theta}\exp\left(-K-\frac{(K\hspace{-1mm}+\hspace{-1mm}1)h}{\Theta}\right)\sum_{k=1}^{4}\alpha_k\exp\left({2\sqrt{\frac{K(K\hspace{-1mm}+\hspace{-1mm}1)h}{\Theta/{\beta_i^2}}}}\right)
   \end{aligned}
   \end{equation} 
 Let us now focus on approximate the PDF of the lognormal shadowing to find an approximate closed form, where the PDF for the Rayleigh/lognormal distributed fading is given by 
 \begin{equation}
f^{(S)}_z(z)=\int \limits_0^{\infty}\frac{2z}{\sigma}\exp\left(-\frac{z^2}{\sigma}\right)\frac{1}{\sqrt{2\pi}\zeta\sigma}\exp\left(-\frac{(\ln \sigma-\mu)^2}{2\zeta^2}\right)d\sigma.
   \end{equation}
Since the instantaneous channel fading power $h$ is given by $h=z^2$, the PDF of channel fading (power) is given by
\begin{equation}
f^{(S)}_h(h)=\int \limits_0^{\infty}\frac{1}{\sigma}\exp\left(-\frac{h}{\sigma}\right)\frac{1}{\sqrt{2\pi}\zeta\sigma}\exp\left(-\frac{(\ln \sigma-\mu)^2}{2\zeta^2}\right)d\sigma
\end{equation}
Again, with the substitution $t=\frac{(\ln \sigma-\mu)}{\sqrt{2}\zeta}$ and exploiting the approximation for Gaussian-Hermite integration , we can express the PDF as the following
\begin{equation}
f^{(R)}_{h}(h)\approx\int \limits_{-\infty}^{\infty}\exp\left(-t^2\right)\Psi(t)dt\approx \sum_{i=1}^{N}\mathcal{W}(i)\Psi(t_i),
\end{equation}
where
$$\Psi(t)= \frac{\exp\left(-\exp\left(-\sqrt{2} t \lambda -\mu \right) h-\sqrt{2} t \lambda -\mu \right)}{\sqrt{\pi }}.$$
Therefore, the PDF of the time-shared shadowed/unshadowed channel fading (power) is given by
\begin{equation}
f_{h}(h)=(1-\mathcal{T})f^{(R)}_{h}(h)+\mathcal{T}f^{(S)}_{h}(h).
\end{equation}
The PDFs of the channel fading (power) gain are drawn for different channel parameter settings in Fig.~1. Higher values of $K$ and shadowing standard deviation make the PDF curves steeper as channel fading gains get more localized around the mean.


\begin{figure*}
  \centering
   \includegraphics[scale=.425]{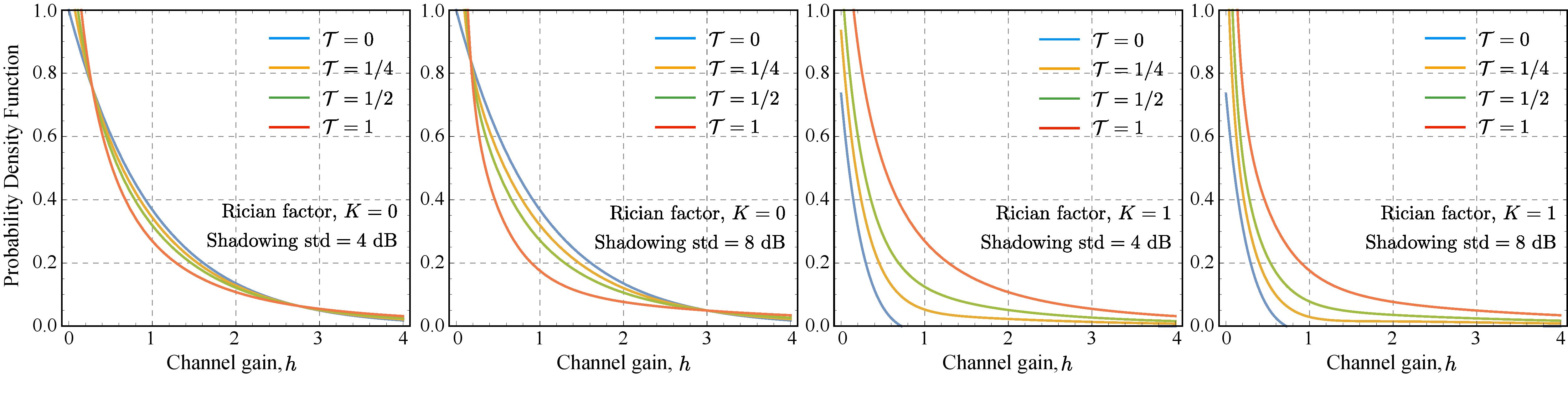}
   \caption{PDFs of time-shared shadowed/unshadowed channel fading (power) for different values of time-sharing factor $\mathcal{T}$ and. Four different plots are drawn for different values of Rician parameter $K$ and shadowing standard deviation. All the curves have been obtained by setting the average power to 1 for both Rician fading and exponential/shadowed fading. In first two plots, the Rician factor $K$ is set to 0, i.e., the case with $\mathcal{T}=0$ mimics the Rayleigh (amplitude) fading.}
   \label{fig1}
\end{figure*}

\section{Outage Analysis for User with and without offloading}
\label{PM}
%
%
The outage probability is an important performance measure of communication links operating over composite fading/shadowing channels. It is defined as the probability that the received SIR falls below a certain protection threshold.
In particular, we are interested in performance metric MU downlink outage probability in direct mode $\mathcal{O}_{\text{d}}^{\text{M}}=1-\mathbb{P}\left[\text{SIR}>\mu\right]$, which is equivalently the CDF of the SIR. On the other hand, the success probability evaluates the complementary CDF (CCDF) of SIR over the entire network. Furthermore, note that we compute the outage probability assuming noise to be zero which mimics the interference limited case, where cumulative interference dominates over noise, typically in cellular radio systems and broadcasting systems where frequency channels are reused in order to achieve higher level of area coverage. Here and for the analysis to follow, we consider the path loss exponent $\eta$ to be 4. This is a widely accepted value for environments where the signal suffers from building obstructions \cite{Hafeez} which aligns with our consideration of the city center as the possible candidate for the deployment of the scheme. 

\subsection{Outage Analysis for Macro Users}
In the general mode of transmission, the MBSs in the MCN transmit to the MUs and there is no inter-tier interference as the MCN and the SCN operate over orthogonal frequencies, i.e., the SCN is not permitted to operate over the spectrum owned by the MCN. The instantaneous SIR of the tagged MU when it is served by MBS can be expressed as

\begin{equation}
{\rm{SIR}}_{o}=\frac{P_{\text{M}}h_mr_m^{-\eta}}{\sum_{j\in\Phi_{\text{M}}\setminus \text{MBS}_o}P_{\text{M}}h_jr_j^{-\eta}},
\end{equation}
where, $\Phi_{\text{M}}\setminus \text{MBS}_o$ is the set of MBSs in $\Phi_{\text{M}}$ except the BS $\text{MBS}_o$ that serves the tagged MU (we assume that the tagged user under consideration is located at the origin $o$). The MU is served by the nearest MBS, $\text{MBS}_o$ located at the random distance $r_m$, and $h_m$ is the channel power fading between $\text{MBS}_o$ and the tagged MU.
Let $\mu_{\rm{M}}$ be the SIR threshold of the MUs. A successful communication between an MU and an MBS is declared when SIR exceeds $\mu_{\rm{M}}$. Therefore, the outage probability $\mathcal{O}_{\text{d}}^{\text{M}}$ is given by 

\begin{equation}
\begin{aligned}
\label{ccdfx}
\mathcal{O}_{\text{d}}^{\text{M}}&=1-p_{\text{s}}^{\text{m}}\mathbb{E}_{r_m}\left[\mathbb{P}[{\rm{SIR}}_{o}>\mu_{\rm{M}}]\hspace{1mm}|\hspace{1mm}{r_m}\right]\\
&=1-p_{\text{s}}^{\text{m}}\mathbb{E}_{r_m}\left[\mathbb{P}\left[h_m>\frac{\mu_{\rm{M}}r_m^{\eta}}{P_{\text{M}}}\left(\sum_{j\in{\Phi_{\text{M}}}\setminus \text{MBS}_o}P_{\text{M}}h_jr_j^{-\eta}\right)\right]\hspace{1mm}|\hspace{1mm}{r_m} \right],\\
&=1-p_{\text{s}}^{\text{m}}\mathbb{E}_{r_m}\left[\mathbb{E}_{I_{\text{M}}}\left[\mathbb{P}\left[h_m>\frac{\mu_{\rm{M}}r_m^{\eta}}{P_{\text{M}}}(I_{\text{M}})\right]\right]\hspace{1mm}|\hspace{1mm}{r_m}\right],\\
\end{aligned}
\end{equation}
where, $r_j$ is the distance of the $j$th interfering MBS from the tagged user captured by $\Phi_{\text{M}}$ and $I_{\text{M}}=\sum_{j\in{\Phi_{\text{M}}}\setminus \text{MBS}_o}P_{\text{M}}h_jr_j^{-\eta}$, and the MBS selects a particular MU that will be served in a designated time slot with probability $p_{\text{s}}^{\text{m}}$.

In the process of finding the outage probability in \eqref{ccdfx}, we can evaluate the inner probability term as
\begin{equation}
\begin{aligned}
\label{ccdfy}
\mathbb{P}\left[h_m>\frac{\mu_{\rm{M}}r_m^{\eta}}{P_{\text{M}}}(I_{\text{M}})\right]&=\hspace{-2mm}\int_{\frac{\mu_{\rm{M}}r_m^{\eta}}{P_{\text{M}}}(I_{\text{M}})}^\infty f_h(h)dh
\end{aligned}
\end{equation}
%
Note that the expectation is to be taken over the interference function.  In the following, in order to keep the analysis tractable and simple, we derive the outage probability for the case of $m=1$ and $\sigma=0$ in \eqref{PDFFH}. We can follow the same procedure to obtain the outage probability for other values of $m$ and $\sigma$ in \eqref{PDFFH} and for time-shared shadowed/unshadowed fading.  Let $f_{I_{\text{M}}}(i)$ be the PDF of the interference. Hence, we can express the outage probability $\mathcal{O}_{\text{d}}^{\text{M}}$ as  
\begin{equation}
\begin{aligned}
\mathcal{O}_{\text{d}}^{\text{M}}&=1-p_{\text{s}}^{\text{m}}\mathbb{E}_{r_m}\left[\int_0^\infty \exp\left(-\frac{\mu_{\rm{M}}r_m^{\eta}}{ P_{\text{M}}}(I_{\text{M}})\right)f_{I_{\text{M}}}(i){\rm{d}}i\hspace{1mm}|\hspace{1mm}{r_m}\right],\\
&=1-p_{\text{s}}^{\text{m}}\mathbb{E}_{r_m}\left[\int_0^\infty \exp\left(-sI_{\text{M}}\right)f_{I_{\text{M}}}(i){\rm{d}}i\hspace{1mm}|\hspace{1mm}{r_m}\right],\\
&=1-p_{\text{s}}^{\text{m}}\mathbb{E}_{r_m}\left[\mathcal{L}_{I_{\text{M}}}[s]|_{s=\frac{\mu_{\rm{M}}r_m^{\eta}}{ P_{\text{M}}}}\hspace{1mm}|\hspace{1mm}{r_m}\right],
\end{aligned}
\end{equation}
where, $\mathcal{L}_{I_{\text{M}}}[s]$ is the Laplace transform of the interference function $f_{I_{\text{M}}}(i)$.
Therefore, outage probability of the MU in direct mode is equal to the Laplace transform of the interference evaluated at $s=\frac{\mu_{\rm{M}}r_m^{\eta}}{ P_{\text{M}}}$. From the definition of Laplace transform and interference expression, we have
\begin{equation}
\mathcal{L}_{I_{\text{M}}}[s]\triangleq\mathbb{E}_{I_{\text{M}}}\left[\exp\left(-sI_{\text{M}}\right)\right]=\mathbb{E}_{\Phi_{\text{M}},h}\left[\exp\left(-s\hspace{-3mm}\sum_{j\in{\Phi_{\text{M}}}\setminus \text{MBS}_o}\hspace{-3mm}P_{\text{M}}h_jr_j^{-\eta}\right)\right].
\end{equation}
It should be noted that the expectation is to be taken over both the PPP and the channel power fading. Due to the independence of the fading coefficients we can express $\mathcal{L}_{I_{\text{M}}}[s]$ as the following
\begin{equation}
\label{pgfl1}
\mathbb{E}_{\Phi_{\text{M}}}\left[\prod_{j\in{\Phi_{\text{M}}}\setminus \text{MBS}_o}\mathbb{E}_{h}\left[\text{exp}\left(-sP_{\text{M}}h_jr_j^{-\eta}\right)\right]\right].
\end{equation}
Using probability generality functional (PGFL) we obtain the Laplace transform of the cumulative interference $I_{\text{M}}$ as
\begin{equation}
\mathcal{L}_{I_{\text{M}}}[s]=\text{exp}\left(-\mathbb{E}_h\left[\int\limits_{r_m}^{\infty}\left(1-\text{exp}\left(-sP_{\text{M}}hr^{-\eta}\right)\right)\lambda_{I_{\text{M}}}(r_m)\text{d}r\right]\right).
\end{equation}
with $\lambda_{I_{\text{M}}}(r_m)=\lambda_{\text{M}}dr^{d-1}\Psi_d$ according to \cite{Haenggi}, where $\Psi_d(r)= |\mathcal{J}(o,r)|=\frac{\pi^{d/2}}{\Gamma(\frac{d}{2}+1)}r^n$ is the volume of $d-$dimensional ball of radius $r$.
Notice that the integration limit ranges from $r_m$ to $\infty$ since the closest interfering MBS is at least at distance $r_m$ from the tagged MU. The interfering MBSs are located within the area $\mathbb{R}^d\setminus \mathcal{J}(0,r_m)$, where $\mathcal{J}(x,y)$ is defined as a ball of radius $y$ centered at point $x$. Therefore, only the MBSs that are outside the ball and transmit on the same RB (or time slot) are interfering the tagged MU.
 Consequently, $\mathcal{L}_{I_{\text{M}}}[s]$ can be expressed as
\begin{equation}
\label{xx1}
\begin{aligned}
\mathcal{L}_{I_{\text{M}}}[s]&\triangleq\exp\left(-\pi \lambda_{\text{M}}\mathbb{E}_h\left(\int_{r_m}^\infty\hspace{-4mm}\left(1-\exp\left(-sP_{\text{M}}hr^{-\eta}\right)\right)d r^{d-1}{\rm{d}}r\right)\right).\\
\end{aligned}
\end{equation}
 Using the substitution $r\leftarrow r^{\delta}$, we can express $\mathcal{L}_{I_{\text{M}}}[s]$ as
\begin{equation}
\begin{aligned}
\mathcal{L}_{I_{\text{M}}}[s]=\exp\left(-\pi \lambda_{\text{M}}\underbrace{\mathbb{E}_h\left[\underbrace{\int_{r_m}^\infty\left(1-\exp\left(-sP_{\text{M}}hr^{-1/\delta}\right)\right){\rm{d}}r}_{a}\right]}_{b}\right),\\
\end{aligned}
\end{equation}
where, $\delta\triangleq d/\eta$. Solving the inner integral $a$ we have 
$$a= r_m\left(\exp\left(-\frac{sP_{\text{M}}h}{r_m^2}\right)-1\right)+\sqrt{\pi sP_{\text{M}}h}Q\left(\frac{\sqrt{sP_{\text{M}}h}}{r_m}\right),$$
where $Q(x)$ is the error function given by $Q(x)=\frac{2}{\sqrt{\pi}}\int_0^x\exp\left(-t^2\right)\text{d}t$. Thereafter, finding the expectation of $a$ when channel fading parameter $h$ has the PDF $f_h(h)$ in \eqref{PDFFH} with $m=1$ and $\sigma=0$, we have

$$b=\left(\sqrt{sP_{\text{M}}}\left(\left(r_m^2+sP_{\text{M}}\right) \text{tan}^{-1}\left[\sqrt{\frac{sP_{\text{M}}}{r_m^2}}\right]\right)\right)/(r_m^2+sP_{\text{M}}).$$

Finally, the outage probability averaged over the plane derived from the Laplace transform of the interference evaluated at $s=\frac{\mu_{\rm{M}}r_m^{\eta}}{ P_{\text{M}}}$ is given by
 \begin{equation}
\begin{aligned}
\mathcal{O}_{\text{d}}^{\text{M}}&=1-\mathbb{E}_{r_m}\left[\mathcal{L}_{I_{\text{M}}}[s]|_{s=\frac{\mu_{\rm{M}}r_m^{\eta}}{ P_{\text{M}}}}\hspace{1mm}|\hspace{1mm}r_m\right],\\
&=1-\int_0^{\infty}\exp\left(-\lambda_{\text{M}} \pi  r_m^2\sqrt{\mu_{\text{M}}} \text{tan}^{-1}\left[ r_m\sqrt{\mu_{\text{M}}}\right]\right) f_{R_{\text{M}}}(r_m)\text{d}r_m.
  \end{aligned}
 \end{equation}
 where $f_{R_{\text{M}}}(r_m)$ is the distance distribution from the tagged MU to the nearest MBS given by
   \begin{equation}
f_{R_{\text{M}}}(r_m)=2\pi\lambda_{\text{M}} r_m\exp\left(-\pi\lambda_{\text{M}} r_m^2\right).
  \end{equation}
  Therefore, the outage probability of the tagged MU in direct mode is given by,
  \begin{equation}
 \mathcal{O}_{\text{d}}^{\text{M}}=1-\int_0^{\infty}\exp\left(-\lambda_{\text{M}} \pi  r_m^2\sqrt{\mu_{\text{M}}} \text{tan}^{-1}\left[ r_m\sqrt{\mu_{\text{M}}}\right]\right) f_{R_{\text{M}}}(r_m)\text{d}r_m.
  \end{equation}
  Similarly, we can obtain the outage probability for the MU served by MBS for time-shared shadowed/unshadowed fading scenarios.

  \subsection{Outage Analysis for Offloaded Macro Users}
  
 In the offloading mode, we consider that all the SBSs in SCN are offloading one MU each during the same time slot.
This creates interference, at levels that depend on the location of the small cells within the MCN coverage area, the transmission power and other environmental factors.
  
 We now investigate the outage characteristics when the tagged MU is offloaded to one of the small cells. The tagged MU is offloaded to the small cell that provides the highest received SIR. In this offloading mode, the MU experiences interference from the the small cells $\Phi_{\text{S}}$ who are providing services to their respective small cell users (SUs) during the same time slot on the shared spectrum. We assume that all the small cells in $\Phi_{\text{S}}$ provide offloading services and during offloading the MBS does not transmit to the tagged MU. The SIR achieved by the tagged MU in the offloading mode can be expressed as
\begin{equation}
{\rm{SIR}}_{o}=\frac{P_\text{S}\max_{i\in\Phi_\text{S}}h_ir_i^{-\eta}}{\sum_{j\in\Phi_{\text{S}}\setminus \text{SBS}_o}P_{\text{S}}h_jr_j^{-\eta}}.
\end{equation}
where, $\Phi_{\text{S}}\setminus \text{SBS}_o$ (i.e., $o=\arg\max_{i\in\Phi_\text{S}}h_ir_i^{-\eta}$) is the set of SBSs in $\Phi_{\text{S}}$ except the BS $\text{SBS}_o$ that serves the tagged SU (we assume that the tagged user under consideration is located at the origin $o$). The MU is served by the nearest SBS, $\text{SBS}_o$ located at the random distance $r_S$, and $H_S$ is the channel power fading between $\text{SBS}_o$ and the tagged MU. A successful offloading, i.e., communication between the tagged MU and an SBS is declared when SIR exceeds $\mu_{\rm{M}}$. Therefore, the outage probability $\mathcal{O}_{\text{o}}^{\text{S}}$ is given by 
  \begin{equation}
\label{ccdfx2}
\mathcal{O}_{\text{o}}^{\text{S}}=1-p_{\text{s}}^{\text{m}}\mathbb{E}_{r_s}\left[\mathbb{E}_{I_{\text{S}}}\left[\mathbb{P}\left[h_s>\frac{\mu_{\rm{M}}r_s^{\eta}}{P_{\text{S}}}(I_{\text{S}})\right]\right]\hspace{1mm}|\hspace{1mm}{r_s}\right]
\end{equation}
where, $r_j$ is the distance of the $j$th interfering MBS from the tagged user captured by $\Phi_{\text{S}}$ and $I_{\text{S}}=\sum_{j\in{\Phi_{\text{S}}}\setminus \text{SBS}_o}P_{\text{S}}h_jr_j^{-\eta}$. 
The PDF of channel power fading coefficients is given by $f_h(h)$
where,  Now, from \eqref{ccdfx}, we can evaluate the inner probability term as
\begin{equation}
\begin{aligned}
\label{ccdfy2}
\mathbb{P}\left[h_s>\frac{\mu_{\rm{M}}r_s^{\eta}}{P_{\text{S}}}(I_{\text{S}})\right]&=\hspace{-2mm}\int_{\frac{\mu_{\rm{M}}r_s^{\eta}}{P_{\text{S}}}(I_{\text{S}})}^\infty f_h(h)\text{d}h
\end{aligned}
\end{equation}

\begin{equation}
\mathcal{O}_{\text{o}}^{\text{S}}=1-p_{\text{s}}^{\text{m}}\mathbb{E}_{r_s}\left[\mathcal{L}_{I_{\text{S}}}[s]|_{s=\frac{\mu_{\rm{M}}r_m^{\eta}}{P_{\text{S}}}}\hspace{1mm}|\hspace{1mm}{r_s}\right],
\end{equation}
where, $\mathcal{L}_{I_{\text{S}}}[s]$ is the Laplace transform of the interference function $f_{I_{\text{S}}}(i)$.
Using PGFL we obtain the Laplace transform of the cumulative interference as
\begin{equation}
\mathcal{L}_{I_{\text{M}}}[s]=\text{exp}\left(-\mathbb{E}_h\left[\int\limits_{r_s}^{\infty}\left(1-\text{exp}\left(-sP_{\text{S}}hr^{-\eta}\right)\right)\lambda_{I_{\text{M}}}(r)\text{d}r\right]\right).
\end{equation}
with $\lambda_{I_{\text{S}}}(r_s)=\lambda_{\text{S}}\Psi_d(r)dr^{d-1}$. 
 Consequently, $\mathcal{L}_{I_{\text{S}}}[s]$ can be expressed as
\begin{equation}
\label{xx1}
\begin{aligned}
\mathcal{L}_{I_{\text{S}}}[s]&\triangleq\exp\left(-\pi \lambda_{\text{S}}\mathbb{E}_h\left(\int_{r_s}^\infty\hspace{-4mm}\left(1-\exp\left(-sP_{\text{S}}Hr^{-\eta}\right)\right)d r^{d-1}{\rm{d}}r\right)\right).\\
\end{aligned}
\end{equation}
%
%

Finally, the outage probability averaged over the plane derived from the Laplace transform of the interference evaluated at $s=\frac{\mu_{\rm{M}}r_s^{\eta}}{ P_{\text{S}}}$ is given by
 \begin{equation}
\begin{aligned}
\mathcal{O}_{\text{o}}^{\text{S}}&=1-\mathbb{E}_{r_s}\left[\mathcal{L}_{I_{\text{S}}}[s]|_{s=\frac{\mu_{\rm{M}}r_s^{\eta}}{ P_{\text{S}}}}\hspace{1mm}|\hspace{1mm}r_s\right],\\
&=1-\int_0^{\infty}\exp\left(-\lambda_{\text{S}} \pi  r_s^2\sqrt{\mu_{\text{M}}} \text{tan}^{-1}\left[ r_s\sqrt{\mu_{\text{M}}}\right]\right) f_{R_{\text{S}}}(r_s)\text{d}r_s.
  \end{aligned}
 \end{equation}

\section{Performance Analysis}
\label{XM}
As we have obtained the expressions for outage probability of the tagged MU both in unoffloaded and offloaded modes, it is now essential to study outage characteristics in different system setups and composite fading scenarios. We have assumed $P_{\text{M}}$ = 43 dBm and $P_{\text{S}}$ = 23 dBm. The MBS density is set to 4 per unit area, and the SBS density is varied to see how the rate coverage performance curves evolves. In order to keep the analysis simple, we have fixed the value $p_\text{s}^{\text{m}}$ to 0.25.
%

\begin{figure}
  \centering
   \includegraphics[scale=.43]{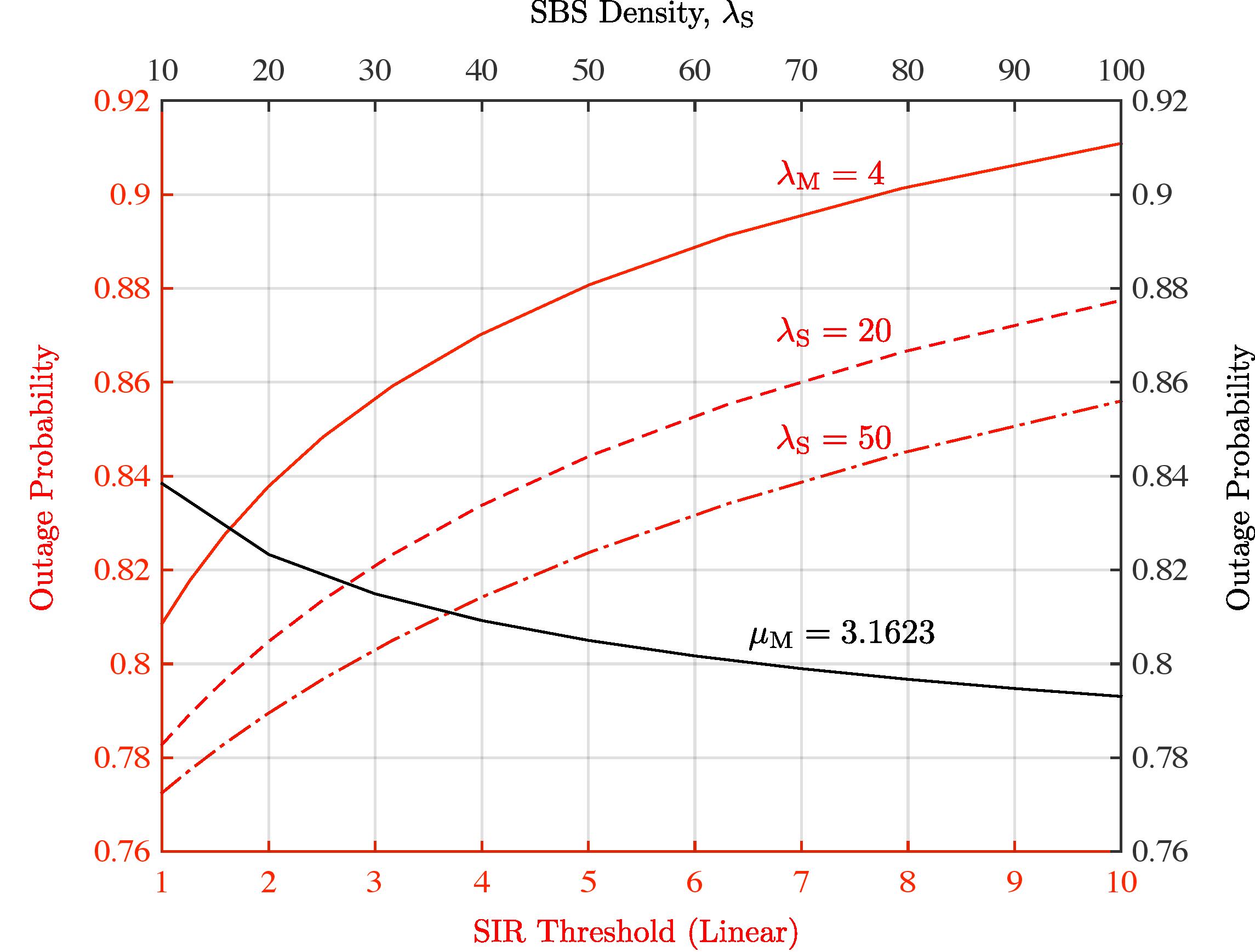}
   \caption{Impact of varying SBS density and SIR threshold on the outage performance of MU with and without offloading.}
   \label{fig1}
\end{figure}
Fig.~\ref{fig1} compares the MU's outage probability with and without offloading services provided by SCN versus SIR thresholds in the case of Nakagami-lognormal channel amplitude fading interfering signals. The Nakagami parameter $m$ and shadowing standard deviation $\zeta$ are set to 1 and 0 (shadowing effect diminishes), respectively, i.e., the channel fading (amplitude) has Rayleigh distribution with unit local mean power. It can be seen that offloading results in a lower outage performance.
With increased SBS density, small cells being placed closer to subscribers, offloading can be more spectrally efficient in a high MCN traffic scenarios.
 As the SIR threshold increases, the outage probability also increases as expected. However, we can observe a significant enhancement in the outage performance as the density of the small cell base stations becomes higher (see the black curve). For example, when a MU in MCN with $\lambda_{\text{M}}=4$ is offloaded to an SCN with $\lambda_{\text{S}}=50$, the outage probability drops down from 0.88 to 0.82 when the SIR threshold is fixed to 5 dB.  This is because as the density of SBSs per unit area increases, the tagged offloading MU comes more closer to the serving SBS.

\begin{figure}
  \centering
   \includegraphics[scale=.43]{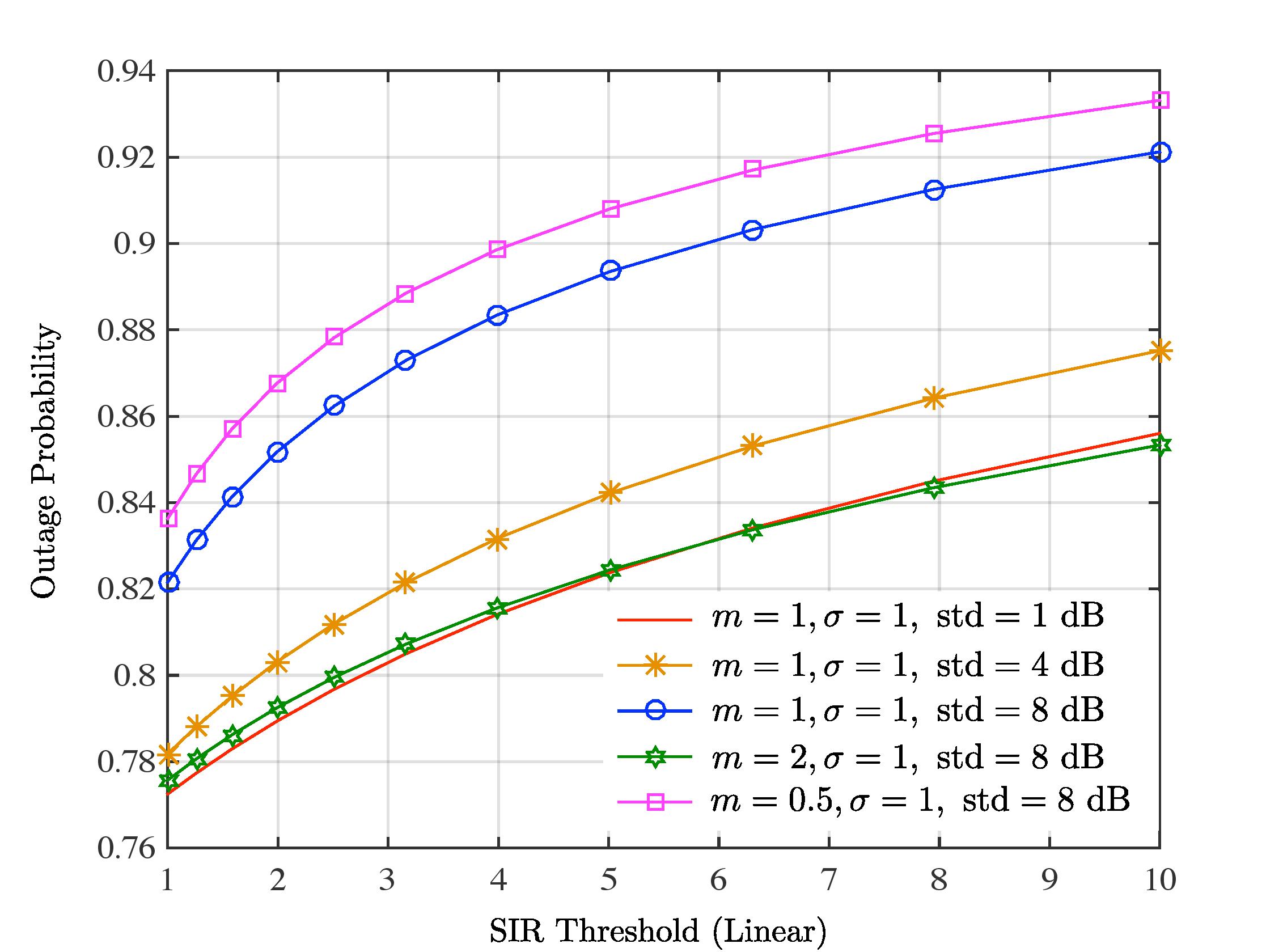}
   \caption{Impact of Nakagami-lognormal shadowing on offloaded MU's outage performance with different fading parameters.}
   \label{fig2}
\end{figure}
In Fig.~\ref{fig2}, we analyze how the Nakagami fading parameter $m$ and lognormal parameters $\mu$ and $\zeta$ impact the outage performance. The SBS density $\lambda_{\text{S}}$ is set to 50 and SIR threshold is varied from 1 dB to 10 dB. It can be seen that the existance of shadowing greatly affect the MU's outage performance when it is offloaded to SCN. For an increase of shadowing standard deviation from 4 dB to 8 dB, we observe a significant deterioration in the SIR coverage.

\begin{figure}
  \centering
   \includegraphics[scale=.43]{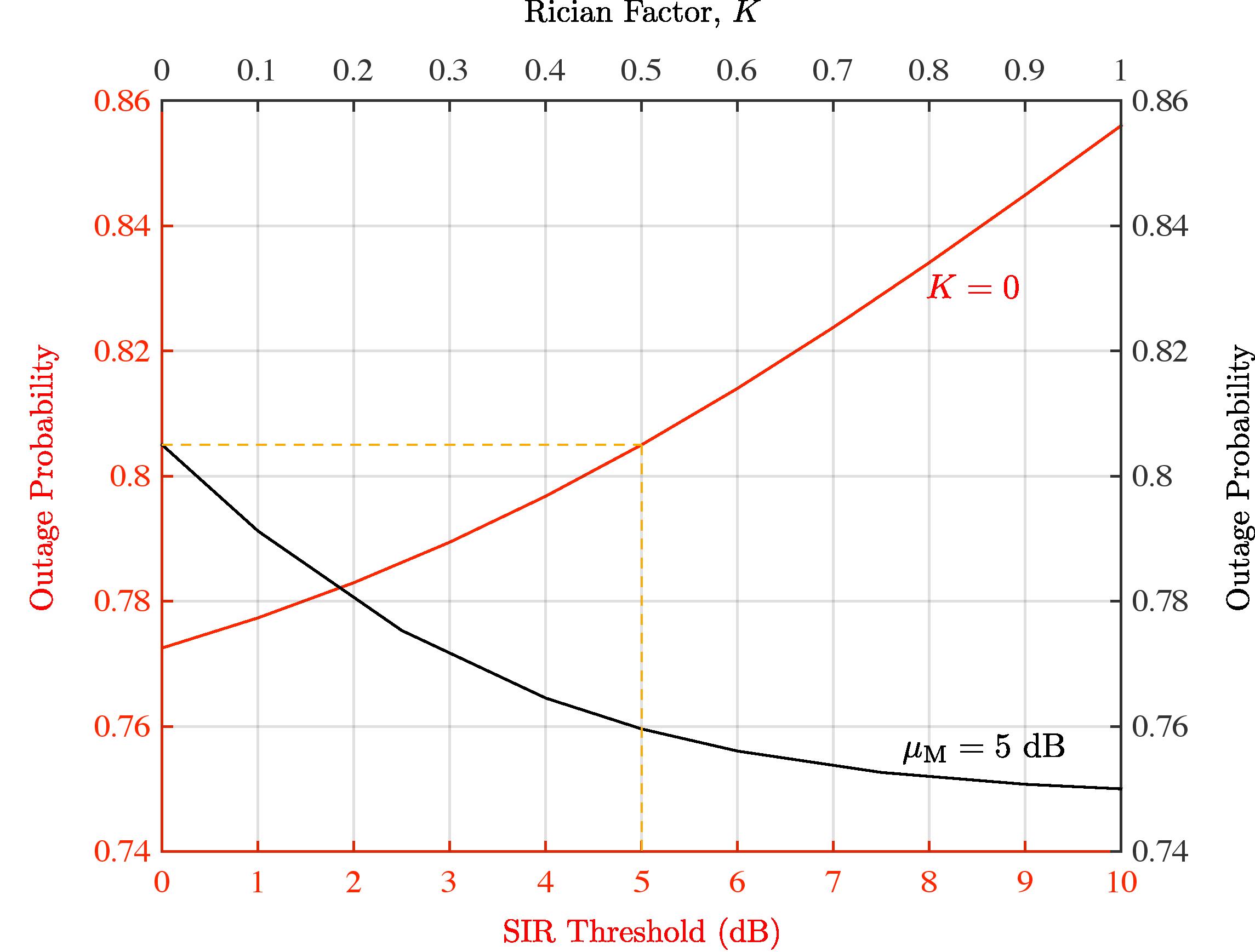}
   \caption{Outage performance of the offloaded MU in Rician fading scenarios.}
   \label{fig3}
\end{figure}
In Fig.~\ref{fig3}, we evaluate the outage performance of the offloaded MU in Rician fading scenarios. When no shadowing is present in time-shared shadowed/unshadowed fading, i.e., $\mathcal{T}=0$ , the channel fading amplitude follows a Rice (Nakagami-$n$) PDF, and the corresponding outage performance is depicted by the red curve with Rician factor $K=0$, therefore, channel fading power is exponentially distributed. Note that as the value of parameter $K$ increases, power in the line-of-sight component increases, the outage probability decreases as expected. The leftmost point of the black curve (i.e., $K=0$) corresponds to the Rayleigh distributed channel amplitude fading and the rightmost point ($K=1$) corresponds to Rice faded channel amplitude fading with equal power in line-of-sight component and multipaths.

\begin{figure}
  \centering
   \includegraphics[scale=.43]{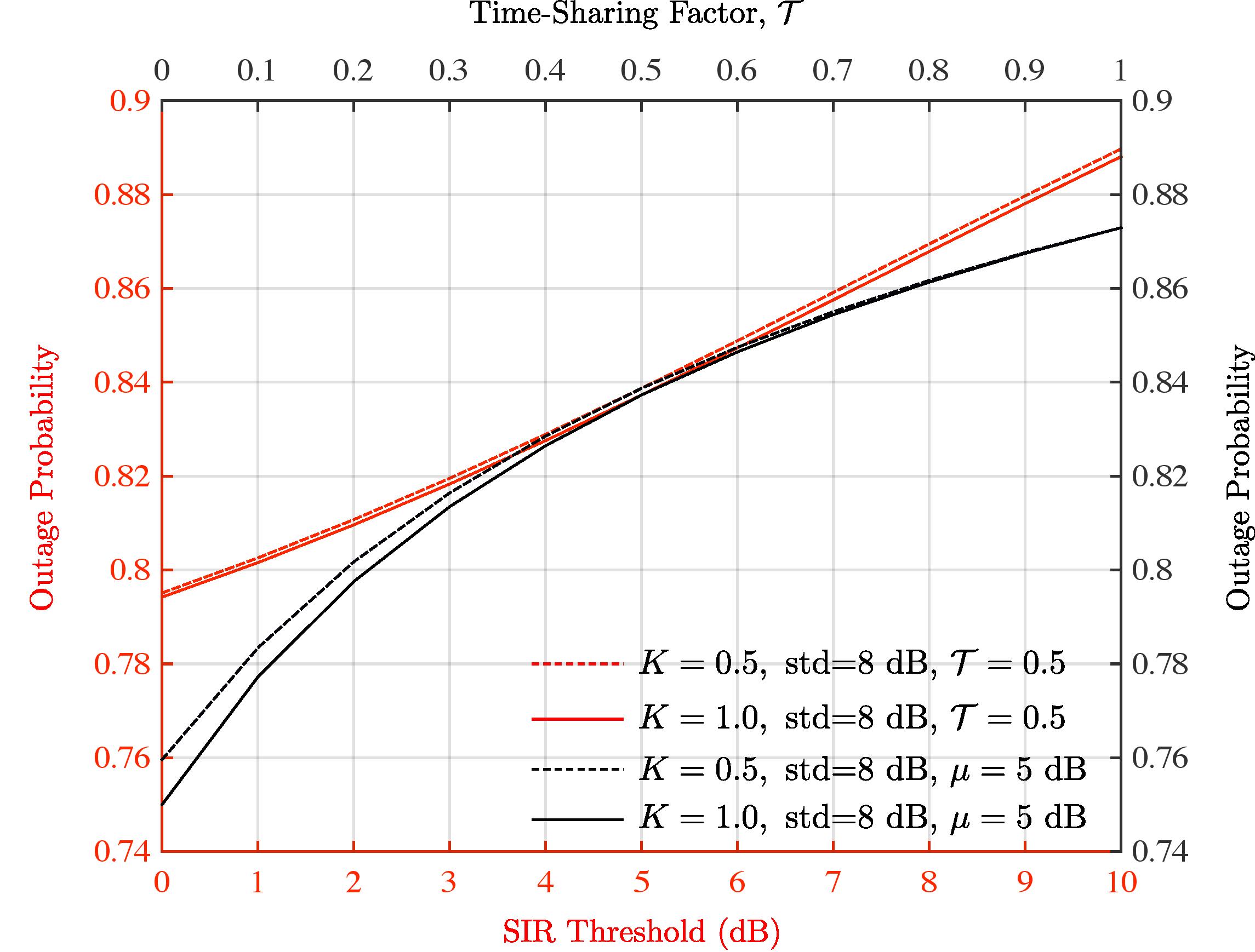}
   \caption{Impact of time-sharing factor $\mathcal{T}$ on the outage performance of the offloaded MU as SIR threshold is varied.}
   \label{fig4}
\end{figure}

In Fig.~\ref{fig4}, we analyze the outage performance of offloaded MU in time-shared shadowed/unshadowed fading scenarios. Recall that overall fading process is a convex combination of unshadowed multipath fading and a composite multipath/shadowed fading. The combination is characterized by the shadowing time-share factor $\mathcal{T}$. The red curve is obtained by setting the time-share factor $\mathcal{T}$ to 0.5, therefore, both the desired and interfering signals experience both Rician fading and Rayleigh-lognormal fading in a time-sharing manner. This situation can be referred to a scenario where for a period equal to half of the slot time the signals experience Rician fading (no shadowing) and for the rest half of the slot time, they experience Rayleigh-lognormal fading (no line-of-sight component). The outage probability increases as SIR threshold increases. The black curve depicts the outage performance when the time-sharing factor $\mathcal{T}$ is varied from 0 to 1 under fixed SIR threshold of 5 dB, therefore, the fading ranges from pure Rician with $K=1$ (the leftmost end) to pure Rayleigh-lognormal (the rightmost end) with $\mu=1$ and $\zeta=8$ dB. Note that the outage probability of the offloaded MU user when the channel fading has Rayleigh-lognormal distribution is higher than that of when the channel fading is Rice distributed. 


\section{Conclusions}
\label{YM}

In this study, we have considered offloading of MUs to SCN under various composite fading scenarios under the framework of stochastic geometry HetNets. In particular, two different types of composite fading scenarios are considered, and we have formulated approximated closed-form expressions for the PDFs of the considered composite fading channels. The impact of small cell base station density on the offloading decision has been analyzed in terms of outage probabilities of the MUs both in direct and offloaded modes. Furthermore, the impacts of the various channel fading parameters along with the Rician-$K$ factor and time-sharing factor have also been investigated.

\section*{Acknowledgement}
This research was conducted under a contract of R\&D for Radio Resource Enhancement, organized by the Ministry of Internal Affairs and Communications, Japan.

\end{document}